\documentclass[12pt,prd,aps,amssymb,amsmath,tightenlines,showpacs,a4paper,nofootinbib]{revtex4}
\def\be{\begin{eqnarray}}
\def\ee{\end{eqnarray}}
\def\bea{\begin{eqnarray}}
\def\eea{\end{eqnarray}}
\def\beax{\begin{eqnarray*}}
\def\eeax{\end{eqnarray*}}
\def\half{\frac{1}{2}}
\def\quarter{\frac{1}{4}}
\def\vf{\varphi}

\newcommand{\il}[1]{$#1$}
\usepackage{color}
\usepackage{graphics}

\begin{document}
\title{Analytic Results for a $PT$-symmetric Optical Structure}

\author{H.~F.~Jones\email{h.f.jones@imperial.ac.uk}}

\affiliation{
Physics Department, Imperial College, London SW7 2BZ, UK}

\begin{abstract}
Propagation of light through media with a complex refractive index in which gain and loss are engineered to be $PT$ symmetric has many remarkable features. In particular the usual unitarity relations are not satisfied, so that the reflection coefficients can be greater than one, and in general are not the same for left or right incidence. Within the class of optical potentials  of the form $v(x)=v_1\cos(2\beta x)+iv_2\sin(2\beta x)$ the case $v_2=v_1$ is of particular interest, as it lies on the boundary of $PT$-symmetry breaking. It has been shown in a recent paper by Lin et al. that in this case one has the property of ``unidirectional invisibility", while for propagation in the other direction there is a greatly enhanced reflection coefficient proportional to $L^2$, where $L$ is the length of the medium in the direction of propagation.

For this potential we show how analytic expressions can be obtained for the various transmission and reflection coefficients, which are expressed in a very succinct form in terms of modified Bessel functions. While our numerical results agree very well with those of Lin et al. we find that the invisibility is not quite exact, in amplitude or phase. As a test of our formulas we show that they identically satisfy a modified version of unitarity appropriate for $PT$-symmetric potentials. We also examine how the enhanced transmission comes about for a wave-packet, as opposed to a plane wave, finding that the enhancement now arises through an increase, of $O(L)$, in the pulse length, rather than the amplitude.
\end{abstract}


\maketitle

\section{Introduction - $PT$ Symmetry and Optics}

The study of quantum mechanical Hamiltonians that are $PT$-symmetric but not Hermitian\cite{BB}-\cite{AMh} has recently found an unexpected application in classical optics\cite{op1}-\cite{op9}, due to the fact that in the paraxial approximation the equation of propagation of an electromagnetic wave in a medium is formally identical to the Schr\"odinger equation, but with different interpretations for the symbols appearing therein. The equation of propagation takes the form
\bea\label{opteq}
i\frac{\partial\psi}{\partial z}=-\left(\frac{\partial^2}{{\partial x}^2}+V(x)\right)\psi,
\eea
where $\psi(x,z)$ represents the envelope function of the amplitude of the electric field, $z$ is a scaled propagation distance, and $V(x)$ is the optical potential, proportional to the variation in the refractive index of the material through which the wave is passing. That is, $V(x)\propto v(x)$, where $n=n_0(1+v(x))$, with $|v|\ll 1$. A complex $v$ corresponds to a complex refractive index, whose imaginary part represents either loss or gain. In principle the loss and gain regions can be carefully configured so that $v$ is $PT$ symmetric, that is $v^*(x)=v(-x)$.

The set-up to which this applies is illustrated in Fig.~1:
\begin{center}
\begin{figure}[h]
\resizebox{!}{1.7in}{\includegraphics{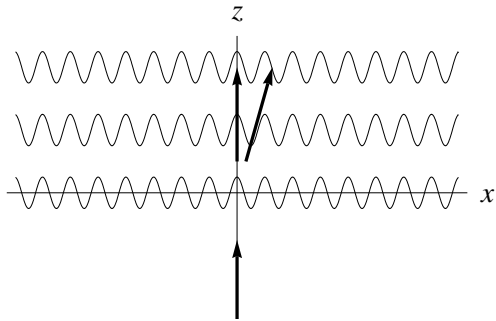}}
\caption{Propagation in the $z$-direction through a material with
refractive index $n=n_0(1+v(x))$. In the paraxial approximation the equation of propagation is formally identical to the time-dependent the Schr\"odinger equation.}
\end{figure}
\end{center}
Most of the initial applications of $PT$-symmetry in optics were concerned with such a set-up, where the variation in $n$ is in the transverse direction. However, recently attention has turned to a different set-up, where the variation in $n$ is in the longitudinal direction, as shown in Fig.~2. In particular the paper by Lin et al.~\cite{Lin} falls into this category. In that case the paraxial equation is not needed, and the scalar Helmholtz equation itself is formally identical to the time-independent Schr\"odinger equation. That is,
\be
\frac{d^2E}{dz^2}+k^2\left(\frac{n}{n_0}\right)^2E=0\ ,
\ee
i.e.
\be\label{Helmholtz}
\frac{d^2E}{dz^2}+k^2(1+2v(z))E=0\ .
\ee
\begin{center}
\begin{figure}[h]
\resizebox{!}{1in}{\includegraphics{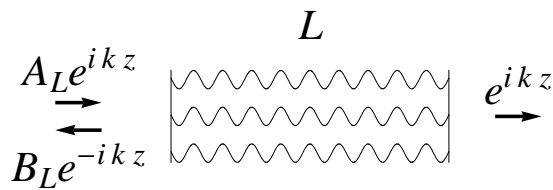}}
\caption{Propagation in the $z$-direction through a slab of medium of length $L$ with refractive index $n=n_0(1+v(z))$. The relevant quantities are now the transmission and reflection coefficients.}
\end{figure}
\end{center}
It is this problem that we wish to discuss in the present paper, for the particular case $v(z)=\half\alpha^2 e^{2i\beta z}$, which was the most interesting case discussed by Lin et al.\cite{Lin}. In the next section we briefly review the approximation scheme used in that paper to calculate the transmission and reflection coefficient. Then in Section 3 we develop an analytic formulation of the problem, whose solution and derived transmission and reflection coefficients can be expressed as simple combinations of modified Bessel functions.  In Section 4 we present our numerical results. Our conclusions are presented in Section 5, which also includes an analysis of the scattering of a wave-packet, as opposed to a plane wave, and a demonstration that the modified unitarity relation developed by Ge et al.~\cite{Ge} is satisfied identically. 


\section{Unidirectional Invisibility}

The paper of Ref.~\cite{Lin} dealt with the general case $v(z)=v_1 \cos(2\beta z)+i v_2 \sin(2\beta z)$, with $v_2$ not necessarily equal to $v_1$. The calculation was performed in a particular approximation scheme, whereby $E$ was written  as
\bea
E={\cal E}_f(z) e^{i\beta z}+{\cal E}_b(z) e^{-i\beta z},
\eea
and second derivatives of ${\cal E}_f$ and ${\cal E}_b$ arising from the Helmholtz equation (\ref{Helmholtz}) were neglected. Moreover, only terms proportional to $e^{\pm i\beta z}$ were kept, while those proportional to $e^{\pm 3i\beta z}$ were dropped, on the grounds that they were more rapidly oscillating. With these approximations, and keeping only linear terms in the detuning $\delta\equiv\beta-k$, one obtains the following coupled linear equations for \il{{\cal E}_f}, \il{{\cal E}_b}:
\be
\frac{d}{dz}\left(\begin{array}{c}{\cal E}_f \cr {\cal E}_b\end{array}\right)=-i\boldsymbol{\sigma.e}\left(\begin{array}{c}{\cal E}_f \cr {\cal E}_b\end{array}\right),
\ee
where $\boldsymbol{e}=\left(- \half v_2 k,\ - \half i v_1 k,\ \delta\right)$,
whose length is $\lambda=\sqrt{\delta^2-\quarter k^2(v_1^2-v_2^2)}$.

These equations are readily integrated, to give the following transmission and reflection coefficients for
left or right incidence:
\bea\label{Kottos}
T_L&=&T_R=\lambda^2/D\cr&&\cr
R_L&=&\left[k^2(v_1-v_2)^2\sin^2(\lambda L)\right]/(4D)\cr&&\cr
R_R&=&\left[k^2(v_1+v_2)^2\sin^2(\lambda L)\right]/(4D),
\eea
where the denominator is $D=\sqrt{\lambda^2\cos^2(\lambda L)+\delta^2\sin^2(\lambda L)}$.

Note that although the transmission coefficients are the same for left or right incidence the reflection coefficients differ markedly. This is because the medium is directional: it is not symmetric under the parity
operation $z\to -z$, but only under the combination $PT$.

The special case when $n_2=n_1$ exhibits the most striking results. In that case $\lambda^2=\delta^2$, so that $D=\lambda$, and Eq.~(\ref{Kottos}) becomes
\bea\label{ideal}
T_L&=&T_R=1\cr&&\cr
R_L&=&0\cr&&\cr
R_R&=&k^2v_1^2\sin^2(\delta L)/\delta\ .
\eea
Thus for left incidence one appears to have perfect transmission and no reflection. It is even the case that within the above approximations the transmission phase is identically zero, which gave rise to the terminology ``unidirectional invisibility". Equally striking is the large enhancement of the reflection coefficient for right incidence, which has the form of the square of a sinc function, whose maximum is proportional to $L$. That is, for right incidence the reflected power grows like $L^2$.


\section{Analytic Solution}
Recall that the analogue Schr\"odinger equation is
\be
\frac{d^2E}{dz^2}+k^2(1+2v(z))E=0,
\ee
where in the case $v_2=v_1$ it is convenient to write $v(z)=\half\alpha^2 e^{2i\beta z}$

Changing variables to $y=(k\alpha/\beta)e^{2i\beta z}$, the equation becomes
\be
y^2\frac{d^2E}{dy^2}+y\frac{dE}{dy}-(y^2+k^2/\beta^2)E=0,
\ee
This is the modified Bessel equation, with solution $E=CI_\nu(y)+DK_\nu(y)$,
where $\nu=k/\beta$. In the language of quantum mechanics ($\psi\equiv E$),
\be
\psi(z)=CI_\nu(y)+DK_\nu(y)
\ee
This has to be matched on to \il{\psi=A_\pm e^{ikz}+B_\pm e^{-ikz}} at \il{z=\pm L/2}

\subsection{Left Incidence}
Referring to Fig.~2, where the transmitted amplitude is normalized to 1 for convenience, we have the initial conditions
\bea\label{Lcond}
\psi(z)&=&1\phantom{ik}=C I_\nu(y_+) +D K_\nu(y_+)\cr&&\cr
\psi'(z)&=&ik\phantom{1}=(CI'_\nu(y_+) +D K'_\nu(y_+))\times (i\beta y_+)
\eea
at $z=L/2$, where where $y_\pm\equiv \nu \alpha\ e^{\pm i\beta L/2}$.
The solution for $C$ and $D$ is
\beax
C&=&y_+K_{\nu+1}(y_+)\cr
D&=&y_+I_{\nu+1}(y_+)\ ,
\eeax
so that
\be
\psi(z)=y_+\left[K_{\nu+1}(y_+)I_\nu(y)+I_{\nu+1}(y_+)K_\nu(y)\right]\ .
\ee
The first of Eqs.~(\ref{Lcond}) is satisfied by virtue of the Wronskian identity\cite{AS}
\be\label{Wronskian}
K_{\nu+1}(y)I_\nu(y)+I_{\nu+1}(y)K_\nu(y)=1/y\ .
\ee
Similarly
\bea
\psi'(z)&=&y_+(i\beta y)\left[K_{\nu+1}(y_+)I'_\nu(y)+I_{\nu+1}(y_+)K'_\nu(y)\right]\cr&&\cr
&=&y_+(i\beta y)\Big\{\left[K_{\nu+1}(y_+)I_{\nu+1}(y)-I_{\nu+1}(y_+)K_{\nu+1}(y)\right]\cr&&\cr
&&\hspace{1.5cm}+\frac{\nu}{y}\left[ K_{\nu+1}(y_+)I_\nu(y)+I_{\nu+1}(y_+)K_\nu(y)\right]\Big\},
\eea
which satisfies the second of Eqs.~(\ref{Lcond}) by virtue of the same identity (recall that $k=\nu\beta$).

At $z=-L/2$ we have to match $\psi$ with $A_L e^{ikL} +B_L e^{-ikL}$.
The general formulas are
\beax
A_L e^{-ik(z-L/2)}&=& \half[\psi(z)+(i/k)\psi'(z)]\cr
&&\cr
B_L e^{ik(z-L/2)}&=& \half[\psi(z)-(i/k)\psi'(z)]\ ,
\eeax
which give, after some algebra,
\bea\label{LT}
A_L&=& (\half \alpha^2 \nu) e^{ikL}[K_{\nu+1}(y_+)I_{\nu-1}(y_-)-I_{\nu+1}(y_+)K_{\nu-1}(y_-)]\cr
&&\cr
B_L&=& (\half \alpha^2 \nu) e^{-ikL}[-K_{\nu+1}(y_+)I_{\nu+1}(y_-)+I_{\nu+1}(y_+)K_{\nu+1}(y_-)]\ .
\eea


\subsection{Right Incidence}
The set-up is shown in Fig.~3, with the transmitted amplitude again normalized to 1.
\begin{center}
\begin{figure}[h]
\resizebox{!}{1in}{\includegraphics{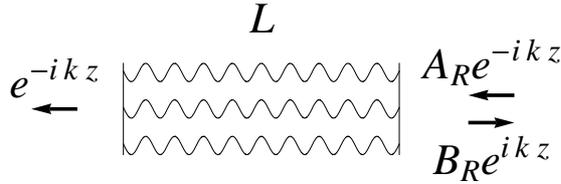}}
\caption{Set-up for propagation from the right}
\end{figure}
\end{center}
The initial conditions at $z=-L/2$ are
\bea\label{Rcond}
\psi(z)&=&1\phantom{-ik}=C I_\nu(y_-) +D K_\nu(y_-)\cr
\psi'(z)&=&-ik\phantom{1}=(CI'_\nu(y_-) +D K'_\nu(y_-))\times (i\beta y_-)\ .
\eea
The solution for $C$ and $D$ is
\beax
C&=&y_-K_{\nu-1}(y_-)\cr
D&=&y_-I_{\nu-1}(y_-),
\eeax
so that
\be
\psi(z)=y_-\left[K_{\nu-1}(y_-)I_\nu(y)+I_{\nu-1}(y_-)K_\nu(y)\right]\ .
\ee
Similarly
\bea
\psi'(z)&=&y_-(i\beta y)\left[K_{\nu-1}(y_-)I'_\nu(y)+I_{\nu-1}(y_-)K'_\nu(y)\right]\cr&&\cr
&=&y_-(i\beta y)\Big\{\left[K_{\nu-1}(y_-)I_{\nu-1}(y)-I_{\nu-1}(y_-)K_{\nu-1}(y)\right]\cr&&\cr
&&\hspace{1.5cm}-\frac{\nu}{y}\left[ K_{\nu-1}(y_-)I_\nu(y)+I_{\nu-1}(y_-)K_\nu(y)\right]\Big\}.
\eea
Again, the initial conditions (\ref{Rcond}) are satisfied by virtue of Eq.~(\ref{Wronskian}).

At $z=L/2$ we have to match $\psi$ with $A_R e^{-ikL} +B_R e^{ikL}$.
The general formulas are
\beax
A_R e^{-ik(z+L/2)}&=& \half[\psi(z)+(i/k)\psi'(z)]\cr
&&\cr
B_R e^{ik(z+L/2)}&=& \half[\psi(z)-(i/k)\psi'(z)]\ ,
\eeax
giving
\bea\label{RT}
A_R&=& (\half \alpha^2 \nu) e^{ikL}[\phantom{-}I_{\nu-1}(y_-)K_{\nu+1}(y_+)-K_{\nu-1}(y_-)I_{\nu+1}(y_+)]\cr
&&\cr
B_R&=& (\half \alpha^2 \nu) e^{-ikL}[-I_{\nu-1}(y_-)K_{\nu-1}(y_+)+K_{\nu-1}(y_-)I_{\nu-1}(y_+)]\ .
\eea

\subsection{Analytic Continuation}
Equations (\ref{LT}) and (\ref{RT}) constitute our main results, but before going on to the numerics a word of caution about the evaluation of the modified Bessel functions is in order. Recall that the argument is $y=(k\alpha/\beta)e^{2i\beta z}$. As $z$ goes from $-L/2$ to $L/2$ this encircles the origin many times, crossing the cut on the negative real axis. Thus we need to know how to continue onto subsequent sheets, which is achieved by using the continuation formulas \cite{AS}
\bea\label{contin}
I_\nu(ye^{im\pi})&=&e^{im\pi}I_\nu(y)\cr
&&\cr
K_\nu(ye^{im\pi})&=&e^{-im\pi}K_\nu(y)-i\pi\ \frac{\sin(m\pi \nu)}{\sin(\pi\nu)}I_\nu(y)\ .
\eea
The resulting functions are smooth functions of $z$, with no discontinuities, as illustrated in the following diagram for $I_{0.6}(y)$.
\begin{center}
\begin{figure}[h]
\resizebox{!}{2in}{\includegraphics{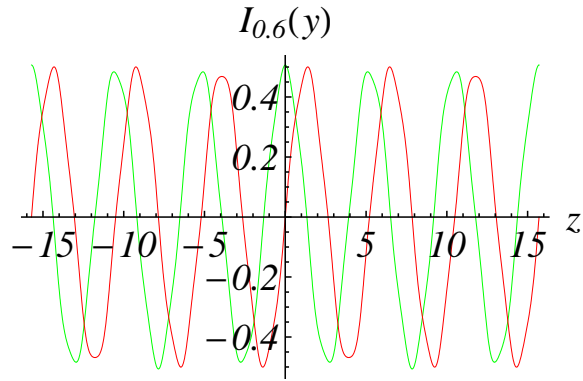}}
\caption{(color online) Real (green) and imaginary (red) parts of $I_{0.6}(y)$ as a function of $z$, continued according to Eqs.~(\ref{contin}).}
\end{figure}
\end{center}
\section{Numerical Results}
The parameters we use in this section are those of Ref.~\cite{Lin}, namely $n_0=1$, $n_1=0.001$, $L=12.5\pi$, $\beta=100$.
\subsection{Left incidence}
For left incidence, we find that $A_L$ is very close to 1, over a wide range of $\delta$, so that the transmission coefficient $T\equiv 1/|A_L|^2$ is very nearly 1, in agreement with Ref.~\cite{Lin}. On closer inspection, however, there is a very small deviation of $Re(A_L)$ from 1 and $Im(A_L)$ from 0 near $\delta=0$, as shown in Fig.~5.
\begin{center}
\begin{figure}[h]
\resizebox{!}{1.7in}{\includegraphics{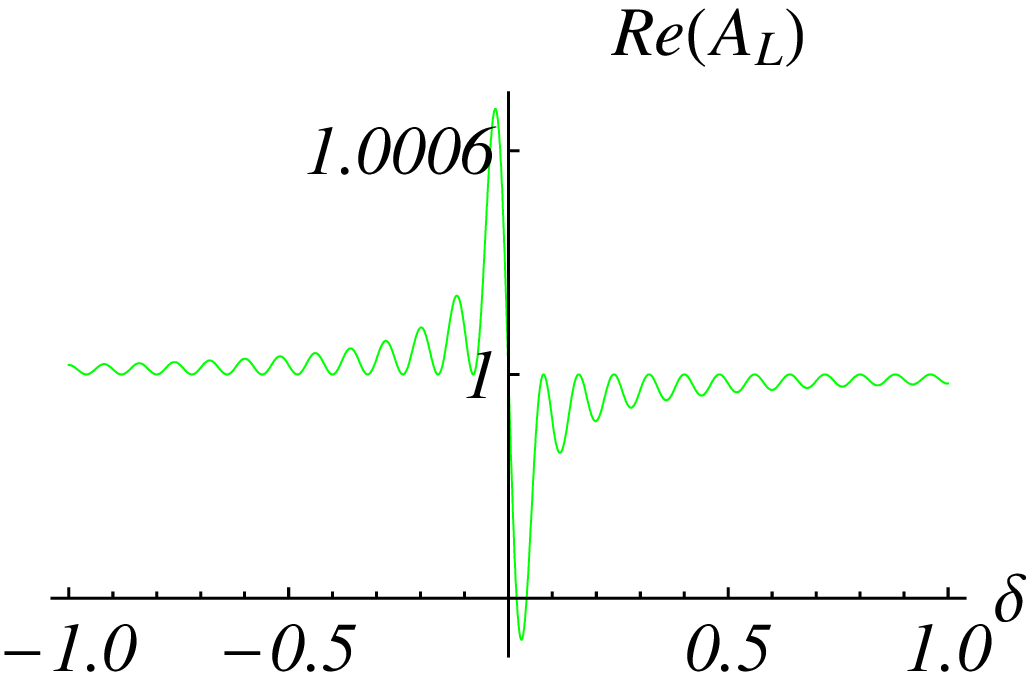}}\hspace{1.5cm}\resizebox{!}{1.7in}{\includegraphics{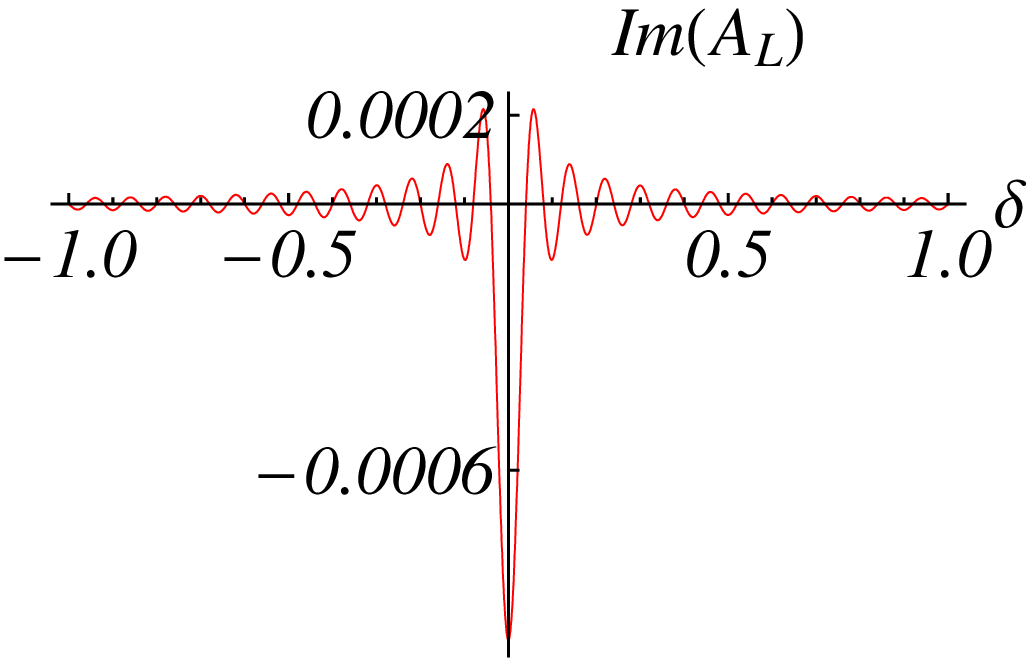}}
\caption{(color on line) Real and imaginary parts of $A_L$ as functions of $\delta$.}
\end{figure}
\end{center}
Correspondingly the transmission is not quite perfect, in amplitude or phase, as shown in Fig.~6.
\begin{center}
\begin{figure}[h]
\resizebox{!}{1.7in}{\includegraphics{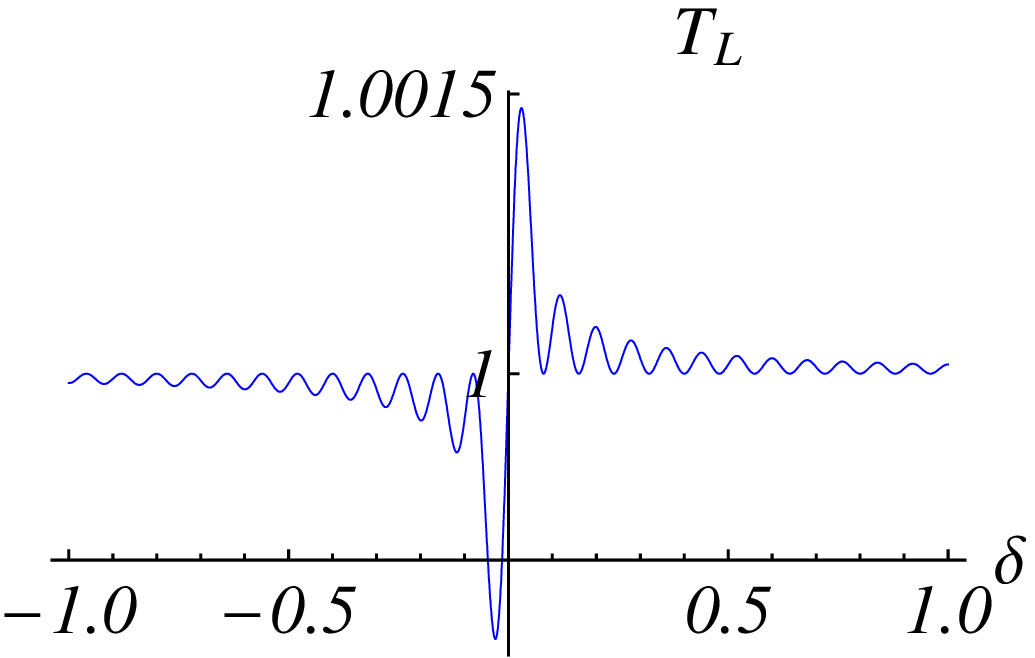}}\hspace{1.5cm}\resizebox{!}{1.5in}{\includegraphics{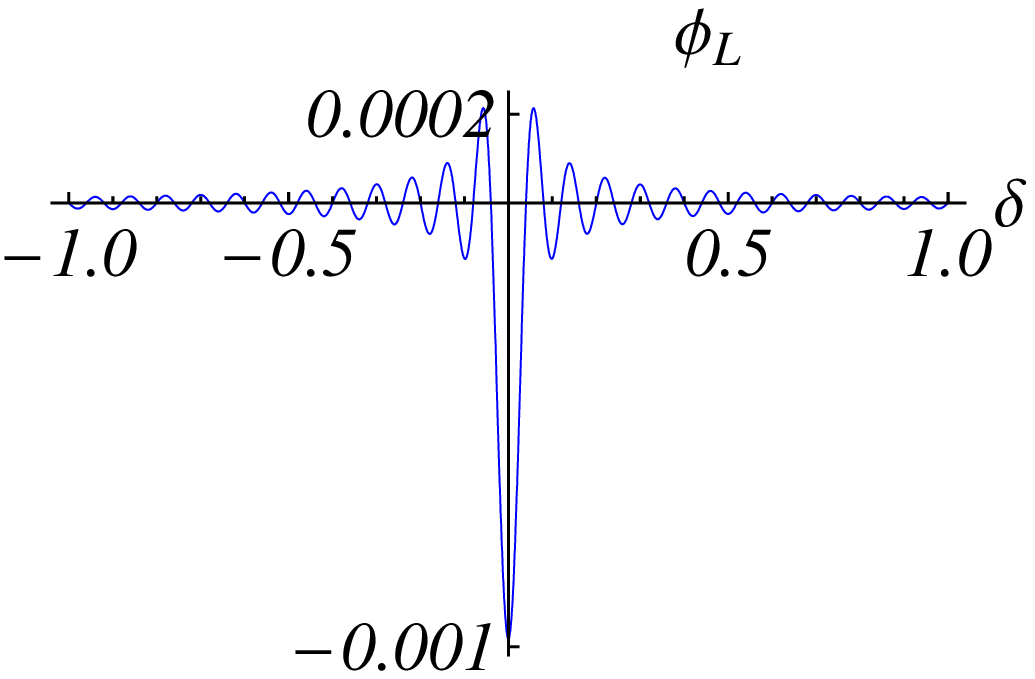}}
\caption{Transmission coefficient and phase of the transmission amplitude as functions of $\delta$.}
\end{figure}
\end{center}
The reflection coefficient $R_L\equiv |B_L|^2/|A_L|^2$ is indeed very small, but varies rapidly on the scale of $10^{-7}$, as shown in Fig.~7.
\begin{center}
\begin{figure}[h]
\resizebox{!}{1.5in}{\includegraphics{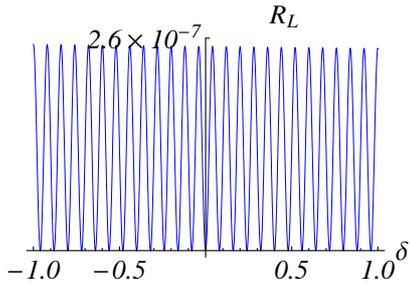}}
\caption{Reflection coefficient $R_L$ as a function of $\delta$.}
\end{figure}
\end{center}
\subsection{Right incidence}
For right incidence the transmission properties are identical, since $A_R\equiv A_L$, but the reflection coefficient differs markedly, as expected from Eq.~(\ref{ideal}). In Fig.~8 we show the real and imaginary parts of $B_R$ and the corresponding reflection coefficient $R_R\equiv |B_R|^2/|A_R|^2$. The latter is almost indistinguishable from the corresponding figure in Ref.~\cite{Lin}.
\begin{center}
\begin{figure}[h]
\resizebox{!}{1.7in}{\includegraphics{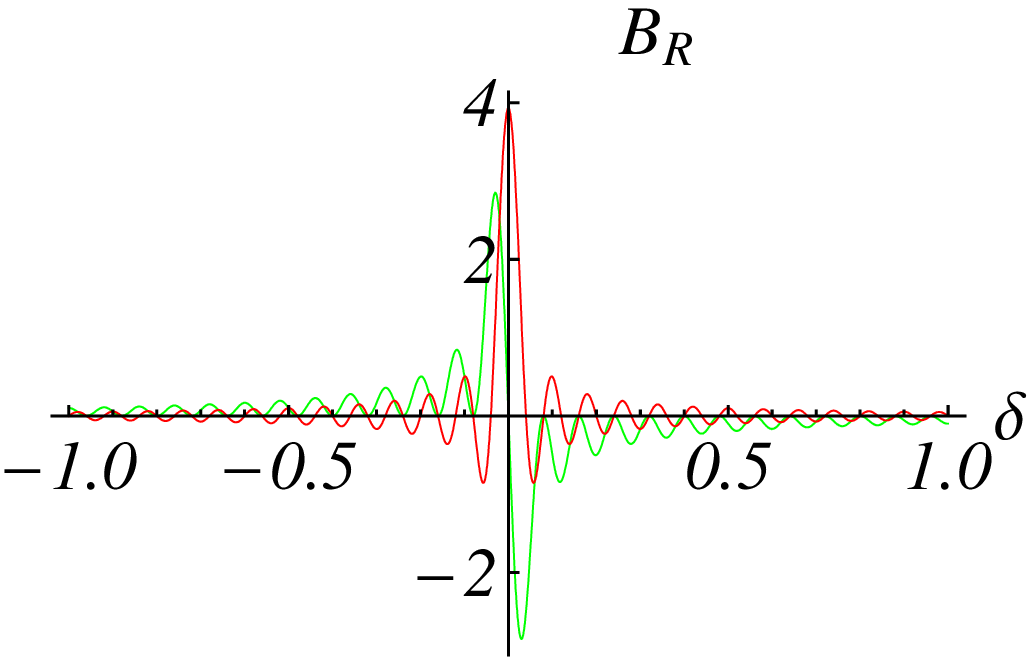}}\hspace{1.5cm}\resizebox{!}{2in}{\includegraphics{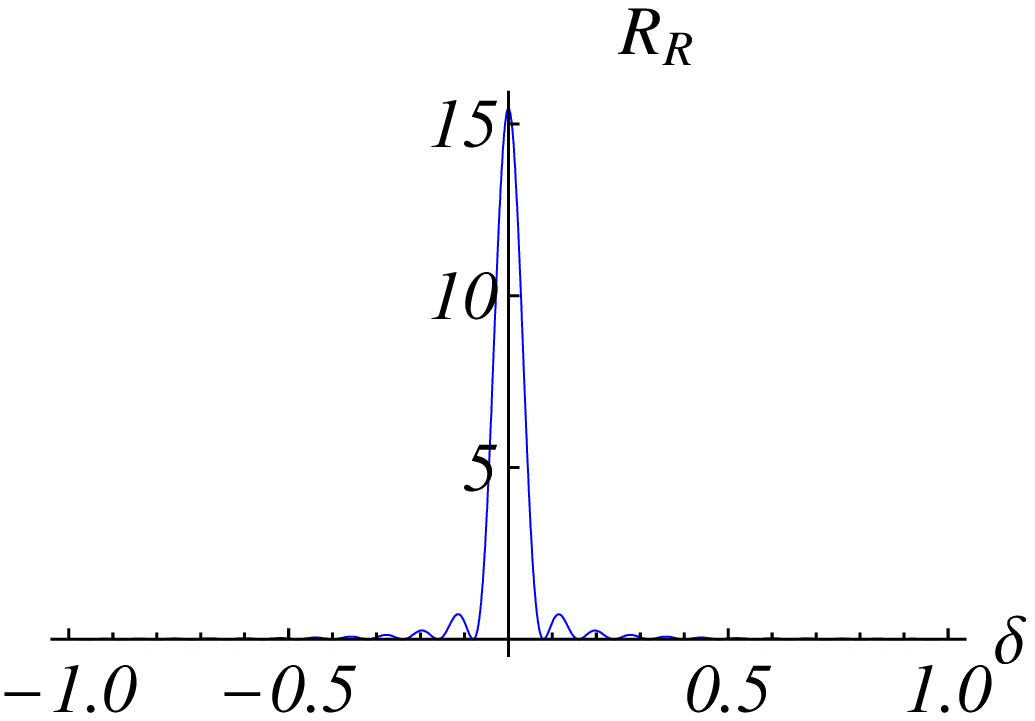}}
\caption{(color on line) Left panel: real (green) and imaginary (red) parts of $B_R$ as functions of $\delta$.
Right panel: reflection coefficient $R_R$ as a function of $\delta$.}
\end{figure}
\end{center}

\section{Discussion}
Our main results are the simple analytic expressions, Eqs.~(\ref{LT}) and (\ref{RT}), for the transmission and reflection amplitudes in the case $v_2=v_1$. It transpires that the approximate expressions given in Ref.~\cite{Lin} are surprisingly good, but that the transmission amplitude $A_{L/R}$ and the reflection amplitude $B_L$ exhibit interesting oscillatory behaviour on a small scale. The large reflection coefficient $R_R$, which is perhaps the most striking effect of the $PT$-symmetric refractive index, is essentially indistinguishable from the simple form of Eq.~(\ref{Kottos}).

A test of our expressions is afforded by the modified unitarity relations recently found by Ge et al.~\cite{Ge} for $PT$-symmetric potentials. Instead of the usual unitary relation $T+R=1$, applicable to a Hermitian potential, one now has the equation
\be
T-1=\pm\sqrt{R_L R_R}\ .
\ee
In the present case, as can be seen from Fig.~(6), the transmission coefficient $T$ is greater/less than 1 for $\delta$ positive/negative.
We can write each side in terms of the amplitudes $A$, $B_{R,L}$ themselves without modulus signs by noting from Eqs.~(\ref{LT}) and (\ref{RT}) that the complex conjugates of $A e^{-ikL}$ and $B_{R,L} e^{-ikL}$  are obtained by interchanging the arguments $y_\pm$. This shows that
$B_R e^{ikL}$ and $B_L e^{ikL}$ are pure imaginary, so that $|B_R B_L| =\pm(B_R e^{ikL})(B_L e^{ikL})^*$ depending on whether $B_R$ and $B_L$ are in or out of phase. After some algebra it can be seen that the modified unitarity relation is satisfied identically by virtue of the Wronskian-type identity
\be
K_{\nu+1}(y) I_{\nu-1}(y)-I_{\nu+1}(y) K_{\nu-1}(y)=\frac{2\nu}{y^2},
\ee
which is not found in Ref.~\cite{AS}, but is readily proved.

In the more general case $v_2<v_1$, an analytic solution is still possible, in terms of Mathieu functions with complex argument, using the similarity transformation \cite{HFJ} that transforms the potential $v_1\cos(2\beta z)+iv_2 \sin(2\beta z)$ into the Hermitian potential $\surd(v_1^2-v_2^2) \cos(2\beta z)$. However, from the mathematical point of view the resulting expressions do not simplify in the manner of Eqs.~(\ref{LT}) and {(\ref{RT}), because of the lack of recursion relations for the Mathieu functions, and from the physical point of view the results are not as dramatic.

Finally it is interesting to investigate how the $L^2$ enhancement of the reflection coefficient is affected when the input is not a continuous plane wave, but rather a wave packet of finite width in $z$ or $t$. The question is of interest because on the one hand it has been shown \cite{Zheng} on general grounds that such an enhancement is a general feature of $PT$-symmetric potentials at the symmetry-breaking point, while on the other hand, in the diffraction set-up of Fig.~1 it has been shown \cite{op6, EMG+HFJ} that for a wave-packet input the maximum amplitude becomes saturated, and the enhanced output is obtained instead from a spreading of the beam.

Including the time dependence, the field $\psi_k(z,t)$ for $z>L/2$ in the case of right incidence is
\bea\label{WP}
\psi_k(z)=A_R e^{-i\left[k(z+L/2)+\omega t\right]}+B_R e^{i\left[k(z+L/2)-\omega t\right]},
\eea
where for $n$ very close to 1 we can take $\omega=k$ in natural units.
In Eq.~(\ref{WP}) we may take $A\approx 1$ and $B_R\approx \half i \alpha^2 k e^{i\delta L}\sin(\delta L)/\delta$ to a very good degree of approximation, in accordance with Eq.~(\ref{ideal}), so that the right-going component is
\be
\psi_\rightarrow(z,t)=\half i \alpha^2 e^{i\beta L} e^{ik(z-L/2-t)}\frac{\sin{\delta L}}{\delta}.
\ee

Then, if we modify the incoming component of the field by folding with a Gaussian shape $(1/(w\sqrt{\pi})e^{-(k-\beta)^2/w^2}$, the left-going component $\vf_\leftarrow(z,t)$ becomes simply a Gaussian proportional to $e^{-\quarter w^2(z+L/2+t)^2}$, while (recall that $k=\beta-\delta$) the right-going component becomes
\bea\label{phi}
\vf_\rightarrow(z,t)&=&\frac{1}{w\sqrt{\pi}}\int_{-\infty}^\infty dk\ e^{-\delta^2/w^2} \psi_\rightarrow(z,t)\cr&&\cr
&=&\frac{i \alpha^2}{2w\sqrt{\pi}}e^{i\beta(z+L/2-t)} \int_{-\infty}^\infty d\delta\ e^{-\delta^2/w^2+i\delta(x-L/2-t)}(\beta-\delta)\frac{\sin{\delta L}}{\delta},
\eea
in which we can neglect $\delta$ in the factor $(\beta-\delta)$ occurring in the integrand, since the large enhancement comes from the $\beta$ term alone.

By using the integral representation $\half \int_{-L}^L d\mu e^{i\delta\mu}$ for $\sin{\delta L}/\delta$ and interchanging the orders of integration we can write
\bea
\int_{-\infty}^\infty d\delta\ e^{-\delta^2/w^2+i\delta \xi}\ \frac{\sin{\delta L}}{\delta}&=&\half w \sqrt{\pi}\int_{-L}^L d\mu\ e^{-\quarter w^2(\xi+\mu)^2}\cr
&=&\half\pi\left[\mbox{erf}\left(\half w(L+\xi)\right)+\mbox{erf}\left(\half w(L-\xi)\right)\right],
\eea
in which $\xi$ is to be identified with $z-L/2-t$. For a wide range of $w$ the two error functions behave like sgn functions, and the combination of the two gives an almost constant plateau for $|\xi|<L$, while essentially vanishing outside that range, as is illustrated in Fig.~(9) for $w=1$ and $L=10$. Thus for a wave-packet input the enhanced reflection comes not from an increase in the overall amplitude, but instead from a lengthening of the pulse, of order $L$. An analogous effect was noted in the diffraction situation by Longhi \cite{op6}, and the connection to the properties of error functions pointed out in Ref.~\cite{HFJ2}.
\begin{center}
\begin{figure}[h]
\resizebox{!}{1.7in}{\includegraphics{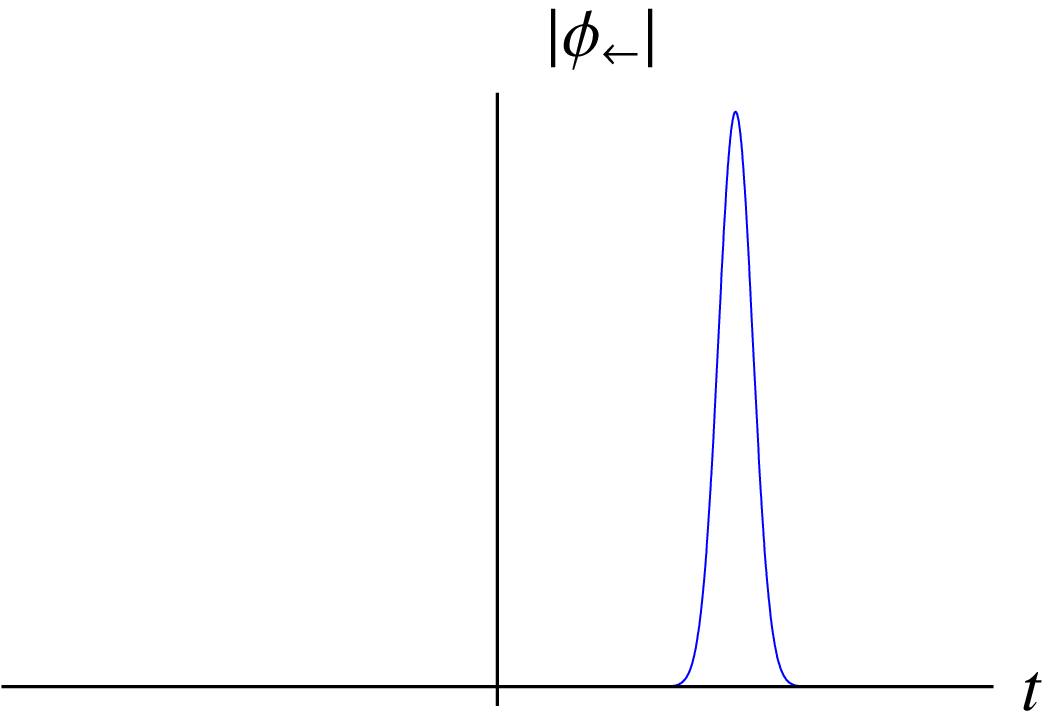}}\hspace{2cm}\resizebox{!}{1.7in}
{\includegraphics{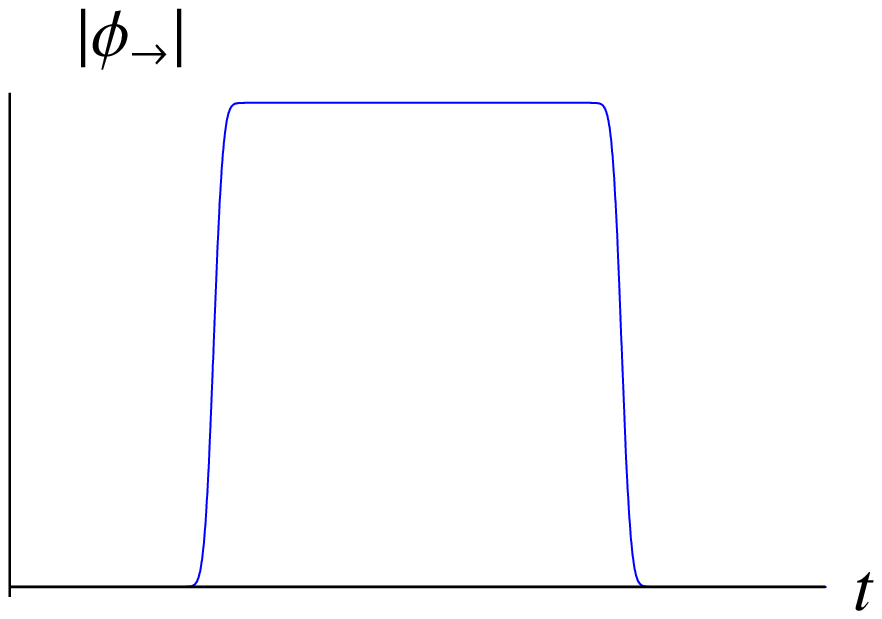}}
\caption{Generic shapes of input and output pulses (left and right panels respectively).}
\end{figure}
\end{center}

\end{document}